\documentclass[pra,twocolumn,preprintnumbers,amsmath,amssymb,showpacs,showkeys]{revtex4}

\usepackage[usenames]{color}
\usepackage{graphicx}
\usepackage{amsfonts}
\usepackage{amssymb}
\usepackage{epsfig}
\usepackage{bm}
\usepackage{dcolumn}
\newcommand{\beq}{\begin{eqnarray}}
\newcommand{\eeq}{\end{eqnarray}}
\newcommand{\e}{{\text e}}

\newcommand{\mf}{\textrm{{\scriptsize MF}}}
\newcommand{\imf}{\textrm{{\scriptsize CMF}}}
\newcommand{\sw}{\textrm{{\scriptsize SW}}}
\newcommand{\lda}{\textrm{{\scriptsize LDA}}}
\newcommand{\In}{\textrm{in}}
\newcommand{\out}{\textrm{out}}
\newcommand{\rin}{\tilde{r}_\In}
\newcommand{\rout}{\tilde{r}_\out}
\newcommand{\gs}{| \textrm{GS} \rangle}
\newcommand{\sg}{\langle  \textrm{GS} |}
\newcommand{\F}{\textrm{F}}

\def\bB{{\mbox{\boldmath $B$}}}
\def\bA{{\mbox{\boldmath $A$}}}
\def\bphi{{\mbox{\boldmath $\phi$}}}
\def\bpsi{{\mbox{\boldmath $\psi$}}}
\def\bk{{\mbox{\boldmath $k$}}}
\def\br{{\mbox{\boldmath $r$}}}

\begin{document}
\title{Analytical and numerical study of trapped strongly \\correlated bosons in two- and three-dimensional lattices}
\date{\today}

\author{Itay Hen}
\affiliation{Department of Physics, Georgetown University, Washington, DC 20057, USA}
\author{Marcos Rigol}
\affiliation{Department of Physics, Georgetown University, Washington, DC 20057, USA}

\begin{abstract}
We study the ground-state properties of trapped inhomogeneous systems of hardcore bosons in two- and 
three-dimensional lattices. We obtain results both numerically, using quantum Monte Carlo techniques,
and via several analytical approximation schemes, such as the Gutzwiller-mean-field approach, 
a cluster-mean-field method, and a spin-wave analysis which takes quantum fluctuations into account. 
We first study the homogeneous case, for which simple analytical expressions are obtained for 
all observables of interest, and compare the results with the numerical ones.
We obtain the equation of state of the system along with other thermodynamic properties such as the
free energy, kinetic energy, superfluid density, condensate density and compressibility. In the 
presence of a trap, there is in general a spatial coexistence of superfluid and insulating domains. We show that the spin-wave-based 
method reproduces the quantum Monte-Carlo results for global as well as for local quantities 
with a high degree of accuracy. We also discuss the validity of the local density approximation. 
Our analysis can be used to describe bosons in optical lattices where the onsite interaction $U$ is 
much larger than the hopping amplitude $t$.
\end{abstract}

\pacs{03.75.Hh,03.75.Lm,67.85.-d,02.70.Ss}

\keywords{superfluidity, Mott insulator, hardcore bosons, local density approximation, XY model}
\maketitle

\section{Introduction}
The successful realization of the superfluid-to-Mott-insulator transition in ultracold bosonic 
gases trapped in optical lattices in one \cite{stoferle04}, two \cite{spielman07,spielman08}, and 
three \cite{greiner02} dimensions has opened the way to a myriad of experimental and theoretical 
studies of strongly correlated lattice systems \cite{bloch08}. One of the main ideas driving these studies 
is that ultracold atoms in optical lattices can be used as analogue simulators of Hamiltonians of 
the Fermi- and Bose-Hubbard type. Intensive efforts are currently under way to validate this approach 
by comparing experimental and theoretical results for systems such as the Bose-Hubbard model which is 
amenable to both treatments \cite{jimenez10,trotzky09}.

In the Bose-Hubbard model, the basic Hamiltonian consists of a hopping term of amplitude $t$ and 
an onsite two-body repulsion term with amplitude $U$. The phase diagram of this homogeneous model 
is known to consist of two phases: (i) a superfluid phase that is present for all 
incommensurate fillings and arbitrary values of the ratio $U/t$, and for commensurate fillings 
below some critical value $(U/t)_c$, which depends on the dimensionality of the system and on the
(integer) filling, and (ii) a Mott insulator which is present for commensurate fillings 
for $U/t>(U/t)_c$ \cite{fisher89,batrouni90,freericks96,kuhner98,sansone08}.

In experiments with ultracold gases, a confining potential is always present. This confining potential 
is to a good approximation harmonic, and generates an 
inhomogeneous density profile in which superfluid and Mott-insulating phases coexist in spatially-separated 
domains \cite{jaksch98,batrouni02,kashurnikov02,kollath04,wessel04,rigol09}. In such systems, the 
appearance of Mott insulating domains in different regions of the trap depends not only on 
the total filling and the ratio $U/t$ (as in the homogeneous case) but also on the curvature 
of the harmonic confining potential. 

So far, an accurate characterization of the local properties of the system in the 
trap has only been achieved by means of quantum Monte Carlo (QMC) simulations 
\cite{batrouni02,kashurnikov02,wessel04,rigol09}, and density-matrix renormalization group 
(DMRG) techniques in one dimension \cite{kollath04}. Our goal in this paper is to introduce an analytical 
approach that provides an accurate prediction of the behavior of the system in two and three dimensions.
In this study we shall consider the strongly correlated limit where the ratio $U/t$ is very large and the 
local occupation of the lattice sites is lower than or equal to one. In that case, the system can  
be described, to a good approximation, by impenetrable (hardcore) bosons. As we will show, this in turn 
enables the implementation of
an analytical treatment which goes beyond simple mean-field calculations, where within this approach, 
quantum fluctuations are taken into account by the addition of spin-wave corrections. 

In what follows, we shall examine the extent to which the results of this method are valid 
by comparing them against quantum Monte Carlo simulations.
As we will show, spin-wave-corrected results provide a big improvement over the usual Gutzwiller-mean-field approach 
and its cluster-mean-field extension. In most cases, for the system sizes considered here, they are also more 
accurate than the QMC-based local density approximation (LDA). In some regimes, and for some local 
observables, the latter is found to be less accurate than the simple Gutzwiller-mean-field approach.

The paper is organized as follows. In Sec. \ref{sec:model} we review the model at hand and discuss 
its properties. In Sec.\ \ref{sec:homo}, we study homogeneous systems in two and three dimensions. 
We obtain the equation of state along with basic thermodynamic properties and examine the extent 
to which several analytical approaches compare against exact numerical QMC simulations. In 
Sec.\ \ref{sec:2dtrap}, we analyze the global and local properties of hardcore bosons trapped in a 
confining harmonic potential in a two-dimensional setup. We also examine the validity of the 
local-density approximation for finite systems and compare the various analytical approximations 
against exact QMC simulations. 
In Sec.\ \ref{sec:3dtrap}, we analyze the behavior of three-dimensional harmonically trapped hardcore bosons 
using the various approximation schemes. Finally, in Sec.\ \ref{sec:conc}, we conclude with a 
discussion and summary of our results.

\section{\label{sec:model}Model}
The Hamiltonian for hardcore bosons confined in a harmonic potential in a $d$-dimensional hypercubic lattice 
with $N=L^d$ sites, can be written as:
\beq
 \label{eq:Ham}
\hat{H} = - t \sum_{\langle ij \rangle} \left( \hat{a}_i^{\dagger} \hat{a}_j 
+ \hat{a}_j^{\dagger} \hat{a}_i \right) -\mu \sum_i \hat{n}_i +V \sum_i r_i^2 \hat{n}_i\,.
\eeq
Here, $\langle ij \rangle$ denotes nearest neighbors and $\hat{a}_i$ ($\hat{a}_i^{\dagger}$) 
destroys (creates) a hardcore boson on site $i$ located at a distance $r_i=|\br_i|$ from the 
center of the trap, where the coordinates $\br_i=(x_{1i}, \ldots,x_{di})$ are given here in units of 
the lattice spacing $a$, which we set to unity. 

The operator $\hat{n}_i=\hat{a}_i^{\dagger} \hat{a}_i$ is the local density operator, $\mu$ is 
the global chemical potential, and $V$ is the harmonic potential strength. The hopping parameter 
$t$ (which we shall fix at $t = 1$) sets the energy scale. The hardcore boson creation and 
annihilation operators satisfy the constraints $\hat{a}^{\dagger 2}_{i}= \hat{a}^2_{i}=0$, 
which prohibit double or higher occupancy of lattice sites, as dictated by the 
$U\rightarrow \infty$ limit of the Bose-Hubbard model. For any two different sites 
$i \neq j$, the creation and annihilation operators obey the usual bosonic relations
$[\hat{a}_{i},\hat{a}_{j}]=[\hat{a}^{\dagger}_{i},\hat{a}^{\dagger}_{j}]=
[\hat{a}_{i},\hat{a}^{\dagger}_{j}]=0$.

To gain a general understanding of the zero-temperature phases of this model, 
let us first analyze the atomic ($t=0$) limit. In this limit, there is no kinetic (hopping) term, 
and the boson number operators $\hat{n}_i$ commute individually with the 
Hamiltonian, so every lattice site is occupied by a fixed number of bosons. 
In this case, the Hamiltonian is diagonal in the Fock-states basis, 
and the ground-state wave-function has the form $\gs = \prod_i^{\otimes} | \psi_i \rangle$ 
and is solved for each site separately, giving
\beq
 | \psi_i \rangle = \left\{
 \begin{tabular}{lll}
$|0 \rangle$  & if & $\mu_i<0$ 
\vspace{0.18cm}\\
$|1 \rangle$  & if & $\mu_i>0$ 
\end{tabular}\right.\,,
\eeq
where we have denoted $\mu_i=\left( \mu -V r_i^2\right)$ as the local chemical potential, and 
$|0\rangle$ and $|1\rangle$ denote vacant and occupied sites, respectively. In the special $V = 0$ 
case, where no trap is present, the model is translationally invariant, and the ground-state 
boson occupancy is the same throughout the lattice: For $\mu < 0$ the minimal energy configuration 
is simply the vacuum, i.e., the completely-empty lattice, and for $\mu > 0$ the minimal energy 
configuration is the completely-filled lattice. The ground-state energy of these phases is 
degenerate at $\mu = 0$. 

In the $t \ne 0$ case, the model has in general no analytic solution \cite{1d}. As it turns out, 
however, the phase boundaries separating the insulating phases from the superfluid one can easily be
obtained analytically in the $V=0$ case even for nonzero $t$. To see this, we use the fact that our 
Hamiltonian commutes with the total-number-of-bosons operator $\hat{N}_b=\sum_i \hat{n}_i$. 
This simply means that for any given $\mu$ and $t$, the ground-state wave-function is a 
linear combination of product states each having the same number of occupied (vacant) sites. 
In the completely-filled phase (which we denote by F) the wave-function 
is simply 
\beq \label{eq:f}
\gs_{\F} = \prod_i^{\otimes} |1\rangle
\eeq
with energy $\varepsilon_{\textrm{F}} = -\mu N$. In the infinitesimally thin layer outside of the F 
phase, the state of the system is characterized by exactly one vacant site, and thus its wave-function 
is the sum
\beq \label{eq:Fdef}
\gs_{\F^{+}} = N^{-1/2} \sum_i\hat{a}_i \gs_F\,,
\eeq
with energy $\varepsilon_{\textrm{F}^{+}} = - 2 d t -\mu (N-1)$. The phase boundary separating the 
superfluid phase and the (insulating) completely filled phase is the curve along which the F state, 
Eq.\ (\ref{eq:f}), is no longer energetically favorable. This happens when its energy becomes 
degenerate with the energy of the defect state, Eq.\ (\ref{eq:Fdef}). Matching the two, we obtain 
the phase boundary:
\beq \label{eq:fbranch}
\frac{\mu}{2 d t } =1 \,.
\eeq
By the same token, the boundary between the superfluid phase and the completely empty insulating phase (which we denote by E) 
is given by $\mu/(2 d t) =-1$. This result can also be obtained by repeating the above exercise with 
the substitution $| 0 \rangle \leftrightarrow | 1 \rangle$. Between the two insulating phases, in the 
region $| \mu/2 d t |<1$, the system is superfluid.

\begin{figure}[htp!]
\includegraphics[angle=0,scale=1,width=0.35\textwidth]{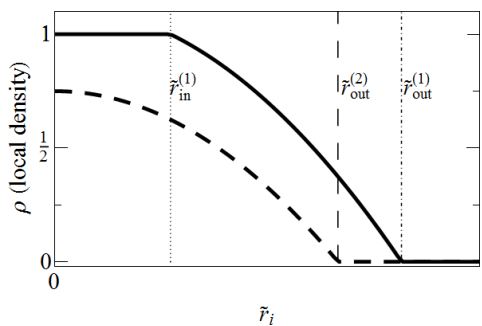}
\caption{\label{fig:schema} Typical density profiles of harmonically trapped hardcore bosons as a function of the distance from
the center of the trap. 
For large enough values of chemical potential (solid line), a Mott insulator with 
$\langle \hat{n}_i \rangle=1$ forms in the center of the trap (the region 
$\tilde{r} \leq r^{(1)}_{\In}$). Outside of this radius the system is superfluid up until 
$\tilde{r}=\tilde{r}^{(1)}_\out$, outside of which the lattice is empty. If the chemical potential 
is small (dashed line) a Mott insulator will not form in the center of the trap, and the only 
superfluid-insulator transition will take place at $\tilde{r}=\tilde{r}^{(2)}_\out$.}
\end{figure}

When a trap is introduced, the system will no longer be entirely superfluid or 
entirely insulating. Depending on the trap curvature and the total chemical potential,
the system may be in a coexistent state of a superfluid in one part of the lattice and an insulator in another. 
Before discussing why this is so, let us first note that unlike the homogeneous case 
which is scale-invariant, in the presence of a trap the model has a length scale determined by the ratio 
of the trap curvature to the hopping amplitude, given by $\xi=(V/t)^{-1/2}$. As a result, our Hamiltonian 
can be rescaled by introducing the dimensionless scaled length  $\tilde{r}_i \equiv r_i/\xi$ through 
which the Hamiltonian can be rewritten in a way that eliminates the amplitude of the harmonic trap $V$:
\beq
 \label{eq:Ham2}
\hat{H} = - t \left[ \sum_{\langle ij \rangle} \left( \hat{a}_i^{\dagger} \hat{a}_j 
+ \hat{a}_j^{\dagger} \hat{a}_i \right) -\sum_i \left( \frac{\mu}{t}- 
\tilde{r}_i^2 \right) \hat{n}_i \right]\,.
\eeq
As a result of this rescaling, it is clear that systems with the same global chemical potential exhibit 
the same qualitative behavior for all positive trap curvatures. In addition, away from the center of the trap, 
the local onsite potential $\mu_i$ always becomes very large and negative and hence outside of some radius $\rout$, 
the lattice must be empty ($\langle \hat{n}_i \rangle =0$). On the other hand, in the center of the trap, 
large enough local chemical potentials will produce an insulating (circular) domain with $\langle \hat{n}_i \rangle =1$ 
inside some critical radius $\tilde{r} \leq \rin$. In the intermediate region, $\rin \leq \tilde{r} \leq \rout$, 
the system is in the so-called superfluid regime. The radial density profiles of the two possible scenarios, 
with and without the insulator in the center of the trap, are sketched in 
Fig.\ \ref{fig:schema}. 

\section{\label{sec:homo}Homogeneous ($V=0$) case in two and three dimensions}

Having discussed the general properties of the model in the previous section, we now turn to analyze 
it in a quantitative manner. We shall try to keep the discussion as general as possible so that it applies
to arbitrary dimensions and confining potentials. Only after the general formulation has been introduced, 
we will obtain analytical expressions for homogeneous (i.e., $V=0$) systems 
in two and three dimensions. (We note that some of the observables of the homogeneous two-dimensional case
we analyze here have been studied in Ref.\ \cite{ber02}.) This general formulation will 
be useful later on when we address the problem of harmonically-confined bosons. 

Our main objective in this section is to obtain the equation of state of the model, i.e., the average 
density as a function of the chemical potential, along with the basic thermodynamic properties of the system. 
Since in dimensions higher than one the model has no analytic solution, exact results for finite systems
(up to the relevant statistical errors) are obtained numerically. Here we use the stochastic series expansion (SSE) 
algorithm \cite{SSE1,SSE2} and perform simulations over a wide range of chemical potentials. 
Zero-temperature properties of the system are obtained by choosing large inverse-temperatures $\beta=1/T$ 
(in our units, $k_B=1$), where we have found it sufficient to have $\beta \geq L$, as the effects 
of increasing $\beta$ beyond this value are indiscernible. In two dimensions, we simulate systems with 
$48 \times 48$ sites and $60 \times 60$ sites, with an inverse-temperature of $\beta=60$, and in three 
dimensions, simulations are performed on a $16 \times 16 \times 16$ lattice and $\beta=20$. After 
obtaining the QMC results, we examine the validity of several approximation schemes 
by comparing them against the results of the SSE simulations.

\subsection{\label{QMDhom}Quantum Monte Carlo results}

\begin{figure}[!b]
\includegraphics[angle=0,scale=1,width=0.48\textwidth]{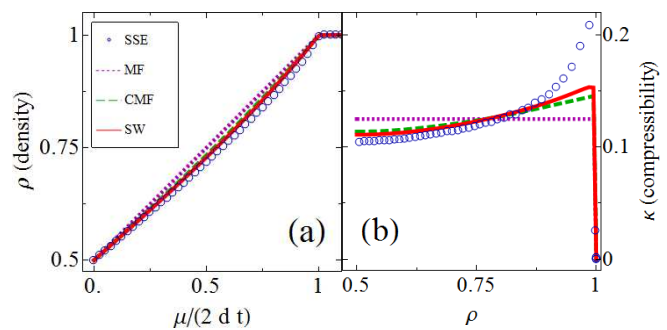}
\caption{\label{fig:rhod2} (Color online) (a) Equation of state and (b) compressibility 
$\kappa=\partial \rho / \partial \mu$ for hardcore bosons in a homogeneous two-dimensional lattice 
($60 \times 60$ sites). The SSE results are denoted by circles, whereas the dotted, dashed, and 
solid lines, are the results of the Gutzwiller-mean-field, the cluster-mean-field, and the spin-wave-corrected 
Gutzwiller-mean-field approximations, respectively.}
\end{figure}

\begin{figure}[!htp]
\includegraphics[angle=0,scale=1,width=0.48\textwidth]{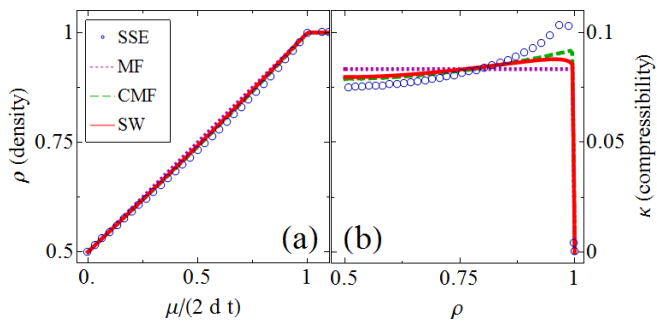}
\caption{\label{fig:rhod3} (Color online) (a) Equation of state and (b) compressibility 
$\kappa=\partial \rho / \partial \mu$ for hardcore bosons in a homogeneous three-dimensional 
lattice ($16 \times 16 \times 16$ sites). The SSE results are denoted by circles, whereas the 
dotted, dashed, and solid lines are the results of the Gutzwiller-mean-field, the cluster-mean-field, 
and the spin-wave-corrected Gutzwiller-mean-field approximations, respectively.}
\end{figure}

We first study the equation of state, which is very relevant for experiments as it provides the dependence of the number 
of bosons in the system (equivalently, the average density) on the global chemical potential. 
It is depicted in Figs.\ \ref{fig:rhod2}(a) and \ref{fig:rhod3}(a) for two- and three-dimensional lattices, respectively. 
The corresponding compressibilities defined by $\kappa = \partial \rho / \partial \mu$ are also given, in 
Figs.\ \ref{fig:rhod2}(b) and \ref{fig:rhod3}(b). These provide the response of the average density to a 
change in the chemical potential. We have computed them in two independent manners: 
(i) as the numerical derivative of the average density with respect to the chemical potential
and (ii) within the SSE algorithm using the formula:
\beq
\kappa = \frac{\partial \rho}{\partial \mu} = \frac1{N Z} \textrm{Tr}\left[ 
\hat{N}_b
\e^{-\beta \hat{H}}
\right]=\frac{\beta}{N} \left( \langle \hat{N}_b^2 \rangle - \langle \hat{N}_b \rangle^2\right) \,,
\eeq
where $Z$ is the partition function and we have taken advantage of the fact that $[\hat{N}_b,\hat{H}]=0$. 
As expected, the results of both methods were found to coincide. In the figures, the SSE results are denoted by circles, 
whereas the various lines indicate the predictions 
obtained by several approximation schemes that will be introduced later. As the figures indicate, at 
$\mu/(2 d t)=0$ the average density is $1/2$. This is because of the particle-hole symmetry of the model. 
Upon increasing the chemical potential, the boson density increases too until it reaches $\rho=1$ at 
$\mu/(2 d t)=1$, as predicted by the analysis given in the previous section. At $\rho=1$ the system becomes insulating and 
the compressibilities drop sharply to zero. Note that due to the particle-hole symmetry, 
the results for negative $\mu$ can be immediately read off the figures by the transformations 
$\mu\rightarrow -\mu$ and $\rho\rightarrow 1-\rho$.

\begin{figure}[!b]
\includegraphics[angle=0,scale=1,width=0.48\textwidth]{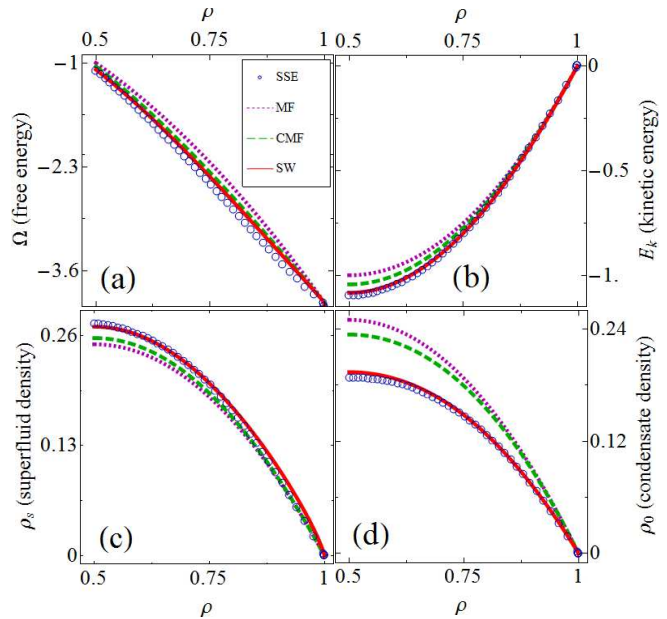}
\caption{\label{fig:alld2homo} (Color online) Thermodynamic properties of hardcore bosons in a 
homogeneous two-dimensional lattice: (a) free energy, (b) kinetic energy, (c) superfluid density, 
and (d) condensate density as a function of the average density. The circles indicate the SSE results 
($60 \times 60$ sites), whereas the dotted, dashed, and solid lines indicate the 
Gutzwiller-mean-field, the cluster-mean-field, and the spin-wave-corrected Gutzwiller-mean-field approximations, 
respectively.}
\end{figure}

\begin{figure}[!ht]
\includegraphics[angle=0,scale=1,width=0.48\textwidth]{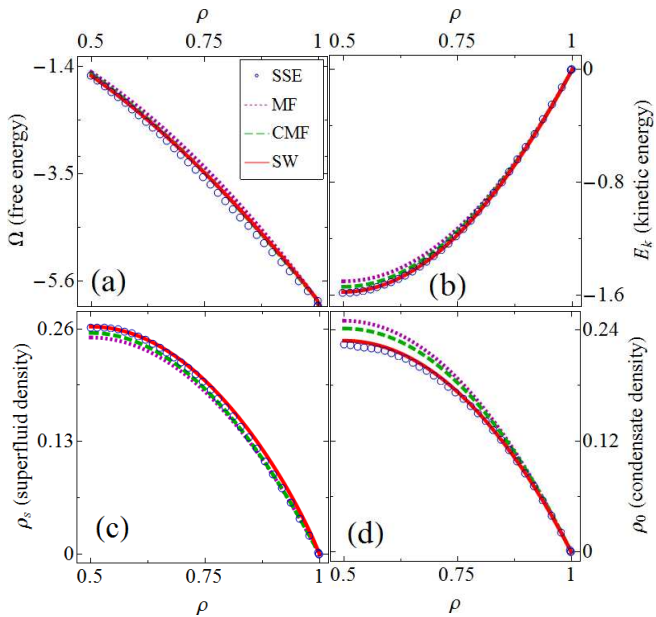}
\caption{\label{fig:alld3homo} (Color online) Thermodynamic properties of hardcore bosons in a 
homogeneous three-dimensional lattice: (a) free energy, (b) kinetic energy, (c) superfluid density, 
and (d) condensate density as a function of the average density. The circles indicate the SSE results 
($16 \times 16 \times 16$ sites), whereas the dotted, dashed, and solid lines indicate the 
Gutzwiller-mean-field, the cluster-mean-field, and the spin-wave-corrected Gutzwiller-mean-field approximations, 
respectively.}
\end{figure}

Figures\ \ref{fig:alld2homo} and \ref{fig:alld3homo} show other observables of interest in 
two and three dimensions, respectively. Those observables provide further understanding of 
the properties of the model: Figures\ \ref{fig:alld2homo}(a) and \ref{fig:alld3homo}(a) show the free energy (per site), 
which is the quantity that is being minimized throughout the simulations. Also shown [Figs.\ \ref{fig:alld2homo}(b) 
and \ref{fig:alld3homo}(b)] is the kinetic energy (per site), which can in principle also be measured in ultracold gases 
experiments by time of flight expansion after simultaneously switching off the trapping potential and the lattice.

The superfluid densities [Figs.\ \ref{fig:alld2homo}(c) and \ref{fig:alld3homo}(c)] and 
the condensate densities [Figs.\ \ref{fig:alld2homo}(d) and \ref{fig:alld3homo}(d)] further indicate 
that for $\rho<1$, when the systems are compressible, they are also superfluid and exhibit 
Bose-Einstein condensation. In the presence of the strong correlations generated by the infinite onsite 
repulsion, one can see that the superfluid density is always greater than the condensate density. 
This is similar to what happens in liquid helium where strong correlations deplete the condensate
density to a very small value while the superfluid density remains very large. Interestingly, as the 
dimensionality of the system increases one can see that the difference between the superfluid density and the 
condensate density decreases. This point will be touched upon later, when we discuss the Gutzwiller-mean-field 
method which corresponds to the exact solution in infinite dimensions. Also evident from the figures is the 
fact that the superfluid density and the condensate density become equal in the low density 
limit.

\subsection{Approximation schemes}

Having analyzed the homogeneous case via the SSE algorithm, we now proceed to analyze the 
model with a number of approximation schemes. We start this investigation with the Gutzwiller mean-field approach.
We note here that this approach is a particular case of a more general treatment initially developed for soft-core bosons \cite{fisher89,jaksch98}.

\subsubsection{\label{sec:mf} Gutzwiller mean-field}

Generalizing the discussion in Refs.\ \cite{hen09} and \cite{hen10} to inhomogeneous systems, 
we start our mean-field calculation with the following product state as our ansatz for the ground state 
wave-function:
\beq \label{eq:GSMF}
\gs_{\mf} =\prod_j^{\otimes} \left( \sin \frac{\theta_j}{2} | 0 \rangle_j + 
\cos \frac{\theta_j}{2} \e^{i \, \varphi_j} | 1 \rangle_j 
\right) \,,
\eeq
where the unknowns $(\theta_j,\varphi_j)$ are to be determined via the minimization of the grand-canonical 
potential (per site) which evaluates to
\beq \label{eq:omegaMF}
\Omega_{\mf} &=\phantom{-}&\frac1{N} \, {_{\mf}} \sg \hat{H} \gs_{\mf} \\\nonumber
&=-&\frac{t}{2 N} \sum_{\langle ij \rangle} \sin \theta_i \sin \theta_j \cos(\phi_i-\phi_j) \\
&\phantom{=}-&\frac1{2 N} \sum_i \mu_i \left(1+\cos \theta_i\right)\,. \nonumber
\eeq
In the homogeneous case, the wave-functions of each of the lattice sites are identical. However, in the 
following, we consider general local potentials, as later we shall apply this method to the case
of the harmonic potential. For the azimuthal angles $\varphi_j$, it is sufficient to require that all 
sites share the same value $\phi_j=\Phi$, where $\Phi$ can be chosen arbitrarily.
The conditions for the polar angles are obtained by differentiating $\Omega_{\mf}$ with respect to 
$\theta_j$, giving:
\beq\label{eq:polCond}
\mu_i \tan \theta_i = t S_i\,.
\eeq
Here, $S_i=\sum_j \sin \theta_j \delta_{i,n(j)}$ where $\delta_{i,n(j)}$ is unity if $i$ and $j$ are 
neighbors and is zero otherwise. This set of equations is to be solved numerically in the general case. 

In the limit where the lattice spacing is very small compared to the average interparticle distance, 
i.e., in the low density limit, the $\theta_j$'s may be assumed to vary smoothly over the lattice and
the above set of equations may be approximated by its continuous version, which turns out to be the 
ordinary differential equation 
\beq\label{eq:ode}
\mu(\br) \tan \theta(\br) = t \nabla^2 [\sin \theta(\br)] + 2 d t \sin \theta(\br)\,,
\eeq
which also has to solved numerically in the general case. 

In the homogeneous case, where $V=0$ and the system is translationally invariant, Eq.\ (\ref{eq:polCond}) 
can be solved exactly, as we can write $\theta_j=\theta$ from which it follows that
\beq
\cos \theta = \textrm{Min}\left[1,\textrm{Max}\left[-1,\frac{\mu}{2 d t}\right]\right] \,.
\eeq

Now that the ground state wave-function has been obtained, all physical properties of interest can easily be
calculated by computing the expectation value of the appropriate operator.
The average density of particles, for example, becomes
\begin{eqnarray}
 \rho_{\mf}&=&\frac1{N} \sum_i {_{\mf}}\sg \hat{a}_i^{\dagger} \hat{a}_i \gs_{\mf}\nonumber\\ &=&
\frac1{2 N}\sum_i \left(1+\cos \theta_i \right)\,,
\end{eqnarray}
where in the homogeneous case, this expression simplifies to
\beq
\rho_{\mf}=\frac1{2}\left(1+\cos \theta \right)\,.
\eeq
The density of bosons is plotted in Figs.\ \ref{fig:rhod2} and \ref{fig:rhod3} for two and three 
dimensions, respectively, along with the corresponding compressibilities, as a function of the chemical 
potential. As can be seen in Figs.\ \ref{fig:rhod2}(a) and \ref{fig:rhod3}(a), when compared against 
the quantum Monte Carlo results, the Gutzwiller mean-field approach provides a good approximation for the density 
close to half filling and again at $\mu/(2 d t)=1$ where the lattice becomes completely filled, but 
deviates from the QMC results in between. The behavior of the compressibility, as predicted by the 
Gutzwiller mean-field approach, is on the other hand qualitatively incorrect for all densities in the
superfluid regime.

Next, the free energy, Eq.\ (\ref{eq:omegaMF}), becomes in the homogeneous case
\begin{eqnarray}
\Omega_{\mf} &=&\frac1{N} \, {_{\mf}}\sg \hat{H} \gs_{\mf}\nonumber\\ &=&- \frac{d t}{2} \sin^2 \theta 
-\frac1{2} \mu(1+ \cos \theta)\,,\label{eq:GSMF2}
\end{eqnarray}
and the density of bosons in the zero-momentum mode (the condensate density) turns out to be
\beq
\rho_{0,\mf}&=& \frac1{N} {_{\mf}}\sg\hat{a}^{\dagger}_{\bk=0} \hat{a}_{\bk=0}  \gs_{\mf}
\nonumber\\
&=&\frac1{4 N^2} \sum_{i,j} \sin \theta_i \sin \theta_j =\frac1{4} \sin^2 \theta \,,
\eeq
where $\hat{a}^{\dagger}_{\bk=0}$ ($\hat{a}_{\bk=0}$) creates (destroys) a particle with momentum $\bk$.
The superfluid density, $\rho_{s,\mf}$, requires a special treatment of the boundary conditions. 
As is well known \cite{rhoS}, the superfluid density is related to the ``spin stiffness'' of the system:
To accomplish this, one needs to compare $\Omega$ (the free energy) of the system under periodic conditions 
with the free energy 
under a ``twist'' in the boundary conditions along one of the linear directions (say, the $x$ direction). 
In the homogeneous case we consider here, the azimuthal angles $\varphi_j$ are all identical.
To implement a twist, we take this angle to be site-dependent and with a constant gradient
such that the total twist across the system in the $x$ direction is $\pi$, namely 
$\delta \varphi = \varphi_{j+\hat{x}}-\varphi_{j}= \pi/L$. 
Within the mean-field treatment, one can show that addition of this gradient 
is equivalent to substituting $t \to t/d \left[ (d-1) + \cos \delta \varphi\right]$.
Now, the square of the gradient twist is related to the superfluid density via the relation 
\hbox{$\Omega_{\textrm{twisted}}-\Omega= t \rho_s  \delta \varphi^2$}
which in turn yields the simple expression
\beq \label{eq:rhos}
\rho_{s,\mf} = -\frac1{2d} \frac{\partial \Omega}{\partial t} \,.
\eeq
At the mean-field level, this expression evaluates to that of the condensate density.
Since mean-field theory becomes exact in infinite dimensions, the differences between both
observables should decrease as the dimension of the systems increases. This explains why the differences 
between the superfluid density and the condensate density (as they were computed with the QMC simulations 
in Sec.\ \ref{QMDhom}) were smaller in three dimensions than they were in two dimensions. 

The quantities discussed in Eqs.\ (\ref{eq:GSMF2})--(\ref{eq:rhos}) are plotted in Figs.\ \ref{fig:alld2homo} and 
\ref{fig:alld3homo} for two and three dimensions, respectively. As one can immediately see, the 
Gutzwiller-mean-field predictions agree rather poorly with the QMC data with respect to all quantities but the free energy,
particularly close to
half-filling. Hence, the mean-field approach described in this 
section does not provide an accurate picture for most observables of interest close 
to half filling. Nonetheless, it is clear that the agreement with the QMC data 
is better in three dimensions: 
As already noted, the mean-field approximation becomes exact in the limit where $d \to \infty$. 

\subsubsection{Cluster mean-field}
The cluster-mean-field approach is an improvement over the simple Gutzwiller-mean-field approach,
which is also applicable to harmonically trapped hardcore bosons. It was introduced 
in Refs.\ \cite{hen09} and \cite{hen10} to improve the Gutzwiller-mean-field prediction for the phase
diagram of hardcore bosons when a period-two superlattice was added to the homogeneous model.
Within this approach, one starts with a variational ansatz which is also a product state, but not
of single-site wave-functions. The new ansatz is a product of wave-functions each describing the state of 
a `block' of $2^d$ sites. In two dimensions, a block consists of $2 \times 2$ square cells, so the ground 
state wave-function has the form:
\beq \label{eq:gsimf}
\gs_{\imf} =\prod_{b}^{\otimes} 
\left(\sum_{i,j,k,l \in \{ |0\rangle,|1\rangle\}} c^b_{ijkl} | i j k l\rangle \right) \,,
\eeq
Analogously, in three dimensions, the basic block is a $2 \times 2 \times 2$ cubic cell \cite{hen09}. 
As with the Gutzwiller-mean-field case, we minimize the free energy 
\hbox{$\Omega_{\imf} = N^{-1} {_{\imf}}\sg \hat{H} \gs_{\imf}$} with respect to the coefficients $c^b_{ijkl}$ 
of the wave-function. Obtaining the various observables in terms of the wave-function given in 
Eq.\ (\ref{eq:gsimf}) is straightforward, and is performed in much the same way as with the usual mean-field 
approach discussed in Sec.\ \ref{sec:mf}. 

The two- and three-dimensional equations of state, as determined by the cluster mean-field approximation, 
are depicted by the dashed lines
in Figs.\ \ref{fig:rhod2} and \ref{fig:rhod3}, respectively. The predictions 
for the other observables are presented in Figs.\ \ref{fig:alld2homo} and \ref{fig:alld3homo}. As the 
figures indicate, the cluster-mean-field predictions are an improvement over the plain Gutzwiller-mean-field 
approach but are still far from accurately reproducing the results of the QMC simulations.

\subsubsection{Addition of spin-wave corrections}
In what follows, we show that the results of the mean-field approaches discussed above can be significantly 
improved by the addition of 
spin-wave corrections which take into account quantum fluctuations \cite{SW1,SW2,SW3,SW4}. This method was
also discussed in the context of homogeneous two-dimensional 
systems in Ref.\ \cite{ber02}. In a later study \cite{hen09}, it was shown that in the presence of a superlattice, 
spin wave corrections do not 
modify the predictions of the simple Gutzwiller-mean-field theory for the phase diagram. Here, we generalize this 
method to arbitrary dimensions and general inhomogeneous systems. 

The spin-wave analysis begins with introducing a set of local rotations to the boson creation and annihilation 
operators. This is accomplished by switching to new field operators via the transformation:
\begin{eqnarray}
\frac1{2}\left( \hat{a}^{\dagger}_j +\hat{a}_j\right) &=&  \frac1{2}\left( \hat{b}^{\dagger}_j +\hat{b}_j\right) 
\cos \theta_j + \left( \hat{b}^{\dagger}_j \hat{b}_j -\frac1{2}\right) \sin \theta_j \,,\nonumber\\
\frac1{2i}\left( \hat{a}^{\dagger}_j -\hat{a}_j\right) &=& \frac1{2i}\left( \hat{b}^{\dagger}_j -\hat{b}_j\right) \,,
\label{eq:trans}\\ \left( \hat{a}^{\dagger}_j \hat{a}_j -\frac1{2}\right)  &=&  \left( \hat{b}^{\dagger}_j \hat{b}_j 
-\frac1{2}\right)  \cos \theta_i - \frac1{2}\left( \hat{b}^{\dagger}_j +\hat{b}_j\right) \sin \theta_j   \,\nonumber.
\end{eqnarray}
The new annihilation and creation operators $\hat{b}_j$ and $\hat{b}_j^{\dagger}$ 
describe low-energy fluctuations about the mean-field ground state -- 
these are spin waves. They too obey hardcore bosons commutation relations. 
Substituting these expressions into our Hamiltonian, ignoring cubic and quartic terms 
in these bosonic operators (thus assuming a dilute gas of spin waves), the new Hamiltonian becomes
a sum of a constant term $\hat{H}_0$, a linear term $\hat{H}_1$ and a quadratic term $\hat{H}_2$, where 
\begin{eqnarray}
\label{eq:H0}
\hat{H}_0 &=&\sum_i \left( -\frac{\mu_i}{2}(1-\cos \theta_i) -\frac1{4} t \sin \theta_i S_i \right) \,,\nonumber\\
\label{eq:H1} 
\hat{H}_1 &=& \sum_i \left( \hat{b}_i^{\dagger} + \hat{b}_i \right) \left( \frac{\mu_i}{2} \sin \theta_i+\frac{t}{2} 
\cos \theta_i S_i  \right)  \,,\nonumber\\
\label{eq:H2} 
\hat{H}_2 &=& \sum_{ij}\left( A_{ij} \hat{b}_i^{\dagger} \hat{b}_j 
+\frac1{2} B_{ij} \left( \hat{b}_i \hat{b}_j +\hat{b}_i^{\dagger} \hat{b}_j^{\dagger}\right) \right) \,,
\end{eqnarray}
and the matrix coefficients in $\hat{H}_2$ are:
\begin{eqnarray}
A_{ij} &=& \left( -\mu_i \cos \theta_i +t \sin \theta_i S_i \right) \delta_{i,j} \nonumber\\
&\phantom{=}&-\frac1{2} t \left( 1+ \cos \theta_i \cos \theta_j \right) \delta_{i,n(j)} \,,\nonumber\\
B_{ij} &=& -\frac1{2} t \left( \cos \theta_i \cos \theta_j -1\right) \delta_{i,n(j)} \,.
\end{eqnarray}

The spin-waves analysis continues with the determination of the $\theta_i$'s via the requirement that the 
linear term $\hat{H}_1$, Eq.\ (\ref{eq:H1}), vanishes. Equating each term in the sum to zero, one ends up with 
the set of equations:
\beq \label{eq:lin0}
\mu_i \tan \theta_i = -t S_i \,,
\eeq
which is analogous to the set of equations obtained previously when the Gutzwiller-mean-field approach was 
discussed, and which similarly has to be solved numerically in the general case. In the homogeneous case where 
$\mu_i=\mu$, we have $\theta_i=\theta$ and therefore also $S_i=2 d \sin \theta$. This leads to the solution:
\beq \label{eq:firstSol}
\cos \theta = \textrm{Min}\left[1,\textrm{Max}\left[-1,-\frac{\mu}{2 d t}\right]\right] \,.
\eeq
Substituting the solution into the zeroth order Hamiltonian $\hat{H}_0$, Eq.\ (\ref{eq:H0}), 
the zeroth order free energy yields, as it should, the free energy of the Gutzwiller mean-field approach:
\beq
\hat{H}_0=-\frac1{2} \sum_i \mu_i +\sum_i \frac{\mu_i \cos \theta_i}{2} \left(1+\frac1{2} \tan^2 \theta_i \right)
\,. 
\eeq
In the homogeneous case, the above expression evaluates to:
\beq
\hat{H}_0/N=- \frac{d t}{2} \sin^2 \theta -\frac1{2} \mu(1+ \cos \theta)\,,
\eeq
which is precisely the mean-field result, Eq.\ (\ref{eq:GSMF2}). 

In order to obtain the spin-wave corrections, one must diagonalize the remaining quadratic term $\hat{H}_2$ in 
Eq.\ (\ref{eq:H2}), for which one assumes that the spin waves are very dilute and hence that the hardcore constraint
can be ignored. The diagonalization process is a lengthy but straightforward process. Here, we shall only 
review the basic steps. The interested reader is referred to Appendix B of Ref.\ \cite{hen09} for a more 
detailed account. The diagonalization process consists of the following steps: (i) Diagonalize the matrix 
$(\bA+ \bB)(\bA-\bB)$ and obtain its (non-negative) eigenvalues $\Lambda^2_k$ along with the unnormalized 
eigenstates $\bphi_k=\{ \phi_{k1},\ldots,\phi_{kl},\ldots,\phi_{kN} \}$ where $k=1 \ldots N$ (here, $\bA$ and 
$\bB$ are matrices whose elements are $A_{ij}$ and $B_{ij}$, respectively). (ii) Evaluate the set of vectors 
$\bpsi_k=\{ \psi_{k1},\ldots,\psi_{kl},\ldots,\psi_{kN} \}$ through the relation $ (\bA-\bB) \bphi_k=\Lambda_k \bpsi_k$\,.  
(iii) Normalize $\bphi_k$ and $\bpsi_k$ according to $\sum_k \bphi_k \cdot \bpsi_k =1$. (iv) Obtain the basic 
two-particle operators through the inverses of $\bphi_k$ and $\bpsi_k$: 
\begin{eqnarray}\label{eq:twoPart}
\langle \hat{b}_k^{\dagger} \hat{b}_{m} \rangle=
\frac1{4} \sum_l \left(  \phi_{kl}^{-1} -  \psi_{kl}^{-1} \right) \left(  \phi_{ml}^{-1} -  \psi_{ml}^{-1}  \right) \,,
\nonumber\\ \langle \hat{b}_k^{\dagger} \hat{b}_{m}^{\dagger} \rangle=
\frac1{4} \sum_l \left(  \phi_{kl}^{-1} -  \psi_{kl}^{-1} \right) \left(  \phi_{ml}^{-1} +  \psi_{ml}^{-1}  \right) \,.
\end{eqnarray}
(v) Express any desired physical quantity in terms of the new field operators using the transformation, 
Eq.\ (\ref{eq:trans}), neglecting cubic and quartic terms, and use the expressions for the two-particle operators, 
Eqs.\ (\ref{eq:twoPart}), to evaluate the resulting expression. 

As an example, consider the local density of particles. Within the spin-wave analysis,  it evaluates to
\beq
\rho_{ii} = \langle \hat{a}_i^{\dagger} \hat{a}_i \rangle = \frac1{2} \left( 1-\cos \theta_i \right) + 
\cos \theta_i \langle \hat{b}_i^{\dagger} \hat{b}_i \rangle \,,
\eeq 
following which the average density becomes:
\beq
\rho_{\sw}&=&  \frac1{N} \sum_i \langle \hat{a}_i^{\dagger} \hat{a}_i \rangle \\\nonumber
&=& \frac1{2N} \sum_i \left( 1-\cos \theta_i \right) + \frac1{N} \sum_i \cos \theta_i 
\langle \hat{b}_i^{\dagger} \hat{b}_i \rangle \,.
\eeq
The potential energy (per site) is then:
\beq
E_{p,\sw}&=& 
 -\frac1{N} \sum_i \mu_i \langle \hat{a}_i^{\dagger} \hat{a}_i \rangle \\\nonumber
 &=& 
-\frac1{2N} \sum_i \mu_i \left( 1-\cos \theta_i \right) - \frac1{N} \sum_i \mu_i \cos \theta_i 
\langle \hat{b}_i^{\dagger} \hat{b}_i \rangle \,.
\eeq
The off-diagonal terms are more cumbersome to calculate but can still be obtained in a straightforward manner. 
They evaluate to:
\begin{widetext}
\beq
\rho_{ij} = \langle \hat{a}_i^{\dagger} \hat{a}_j \rangle &=&
\frac1{4} \left[ \sin \theta_i \sin \theta_j 
\left(1-2  \langle \hat{b}_i^{\dagger} \hat{b}_{i}^{\dagger} \rangle-2 \langle \hat{b}_j^{\dagger} 
\hat{b}_{j}^{\dagger} \rangle \right) \right]
+\frac1{4} \langle \hat{b}_i^{\dagger} \hat{b}_{j} \rangle \left(1+\cos \theta_i +\cos \theta_j +
\cos \theta_i \cos \theta_j \right)  \nonumber\\
&\phantom{=}&+\frac1{4} \langle \hat{b}_j^{\dagger} \hat{b}_{i} \rangle \left(1-\cos \theta_i -
\cos \theta_j +\cos \theta_i \cos \theta_j \right)  
+\frac1{4} \langle \hat{b}_i \hat{b}_{j} \rangle \left(-1+\cos \theta_i -\cos \theta_j +
\cos \theta_i \cos \theta_j \right)  \nonumber\\
&\phantom{=}&+\frac1{4} \langle \hat{b}_i^{\dagger} \hat{b}_{j}^{\dagger} \rangle \left(-1-
\cos \theta_i +\cos \theta_j +\cos \theta_i \cos \theta_j \right)  \nonumber\\
\eeq 
\end{widetext}
Once these are calculated, one can obtain the kinetic energy (per site):
\beq
E_{k,\sw}=&-&\frac{t}{4N} \sum_i \sin \theta_i S_i 
+ \frac{t}{N} \sum_i \sin \theta_i S_i \langle \hat{b}_i^{\dagger}\hat{b}_i \rangle \nonumber\\
&-&\frac{t}{2N} \sum_{ij} \left( 1+ \cos \theta_i \cos \theta_j \right)\delta_{i,n(j)} 
\langle \hat{b}_i^{\dagger} \hat{b}_j \rangle \,, \nonumber\\
\eeq
and also the momentum distribution function defined by:
\beq
n(\bk) = \frac1{N} \sum_{lm} \rho_{lm} \,\e^{- i \bk \cdot (\br_l-\br_m) }  \,.
\eeq

In the homogeneous case, the above equations simplify substantially, and one can obtain analytic 
expressions for the various physical observables \cite{hen09}. The spin-wave corrected density of particles 
is given by:
\beq
\rho_{\sw}= \rho_{\mf} -\frac1{2 N} \cos \theta \sum_{ k \neq 0} 
\left( \frac{\alpha_k}{\sqrt{\alpha_k^2-\beta_k^2}}-1 \right)\,,
\eeq
where we have defined 
\begin{eqnarray}
\alpha_k&=&-\frac{t}{2} \left[ \left(1+\cos^2\theta\right) \gamma_k -2 d \right] \,,\nonumber\\
\beta_k&=&\frac{t}{2} \sin^2 \theta \gamma_k \,,\nonumber\\
\gamma_k&=&\sum_{i=1}^{d} \cos k_d \,.
\end{eqnarray}
The spin-wave corrected density is plotted in Figs.\ \ref{fig:rhod2} and\ \ref{fig:rhod3} for the two- and three-dimensional cases, 
respectively. As is evident from the two figures, the spin-wave corrected densities provide a decisive 
improvement over the plain Gutzwiller-mean-field results and they are also closer to the exact 
results than the cluster mean-field calculations for the most part. Interestingly, we find that for the compressibility, the results of the 
cluster mean-field approach are slightly better than those provided by the spin-wave corrections
in three dimensions. 
 
The spin-wave corrections also yield the following expression for the free energy:
\beq
\Omega_{\sw}= \Omega_{\mf} +\sum_k \left( \sqrt{\alpha_k^2 -\beta_k^2} -\alpha_k \right) \,,
\eeq
where the superfluid density is obtained by evaluating the expression $ \rho_{s,\sw} = -(2 d)^{-1} \, 
\partial \Omega_{\sw}/\partial t$. Finally, the density of bosons in the zero-momentum mode is found to be:
\beq
\rho_{0,\sw}=\rho_{0,\mf} -\frac1{2 N} \sin^2 \theta   \sum_{ k \neq 0} \left( \frac{\alpha_k}
{\sqrt{\alpha_k^2-\beta_k^2}}-1 \right)
\,.\nonumber\\
\eeq
The above quantities are plotted in Figs.\ \ref{fig:alld2homo} (two dimensions) and \ref{fig:alld3homo} 
(three dimensions). As is clear from the figures, not only are the spin-wave corrected predictions a major 
improvement over the mean-field approaches for the thermodynamic quantities, but they are essentially on top 
of the QMC results for the kinetic energy and the superfluid density, and very close to the QMC results 
for the free energy and the condensate density. 

\section{\label{sec:2dtrap}Trapped hardcore boson in two dimensions}

Having discussed the properties of the homogeneous case, we are now ready to analyze the more experimentally-relevant 
case of lattice bosons in the presence of a harmonic trap. We shall limit the discussion to the 
two-dimensional case in this section and address the three-dimensional case in the next section.

As discussed in Sec. \ref{sec:model}, in the presence of a harmonic potential, the trapping amplitude $V$ 
can be scaled out by using the dimensionless parameter $\xi =(V/t)^{-1/2}$. In order to 
verify that this is indeed the case, we have performed simulations over a range 
of chemical potentials for two different trapping amplitudes, $V=0.01$ and $V=0.02$. We have verified that 
once properly normalized, results for local quantities for the two trapping potentials are virtually the same. 

In much the same way we analyzed the homogeneous case in the previous section, 
here we obtain the equation of state of the model along with the various 
thermodynamic properties in the trapped system. We examine the accuracy of the analytical approaches 
discussed in the previous section in this inhomogeneous setup and compare them against QMC simulations in a trap.
(For the trapped bosons, however, the equations derived in the previous section cannot be reduced to simple 
analytical expressions as in the homogeneous case; they must be computed numerically.) In addition, 
we explore the regimes of validity of another approximate method, namely, the LDA which is based on the QMC 
results of the homogeneous system.

\subsection{Equation of state}

In a trapped system, where local densities change with position, the meaning of `equation of state' 
may become somewhat unclear. This is because in an inhomogeneous trapped system 
the local phases are determined not only by the onsite interaction and total filling of the system but 
also through the strength of the confining potential. Fortunately, it turns out that the total filling and 
the curvature of the confining potential come into play only as a combination of the two: This is the characteristic density. 
It is the trap-equivalent of the overall density in homogeneous systems. 

The characteristic density ($\tilde{\rho}$) was introduced for lattice fermions in one dimension in 
Refs.\ \cite{rigol03,rigol04}, where it was shown by means of QMC simulations, that the scaled dimensionless variable 
$\tilde{\rho}=\rho N (a/\xi)^d$ can be used to generate a state diagram of the coexisting phases in the trap in the plane 
$\tilde{\rho}$ vs $U/t$ in a way that is independent of the specific values of the number of particles 
and the amplitude of the confining potential. Further studies validated this approach for the 
one-dimensional fermionic case within Bethe-Ansatz schemes \cite{liu05,campo05,heiselberg06,xianlong06}, 
and state diagrams have been obtained for fermions in higher dimensions by means of dynamical mean-field 
theory (DMFT) \cite{leo08}. The same idea has been shown to work with equal success
in bosonic systems \cite{rigol09}, in the framework of which the concept of characteristic density has been further 
discussed within the local density approximation \cite{batrouni08} and finite-size scaling \cite{camp1,camp2}.
It should be noted however that the concept 
of characteristic density remains a meaningful quantity even in cases where the LDA fails \cite{rigol09}.

\begin{figure}[!htp]
\includegraphics[angle=0,scale=1,width=0.48\textwidth]{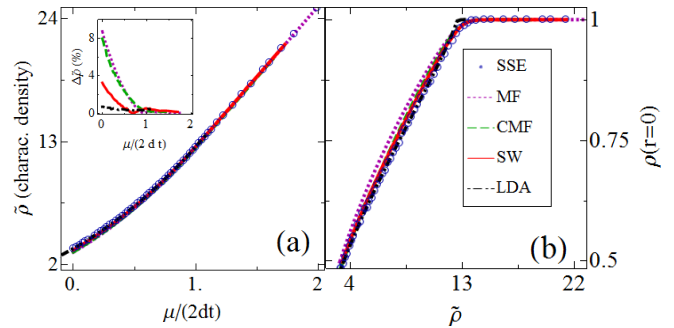}
\caption{\label{fig:eosTrap2d} (Color online) (a) Equation of state and (b) density of bosons in the 
center of the trap vs $\tilde{\rho}$ for harmonically trapped hardcore bosons in a two dimensional lattice 
($48 \times 48$ sites and $V=0.02$). The circles indicate the QMC results whereas the dotted, 
dashed, solid, and dot-dashed lines are the Gutzwiller-mean-field, the cluster-mean-field, the Gutzwiller-mean-field
with spin-wave corrections, and the LDA results, respectively. The inset in panel (a) shows the deviations of 
the various approximation schemes from the SSE data.}
\end{figure}

As follows from the above discussion, one can then generate equations of state for trapped systems
based on the characteristic density. This has recently been done by Roscilde in Ref.\ \cite{roscilde10}
for soft-core bosons in various superlattice potentials. The equation of state for two-dimensional hardcore bosons
is given in Fig.\ \ref{fig:eosTrap2d}(a), which shows the dependence of the characteristic density in our two-dimensional 
trapped systems in terms of the global chemical potential. The SSE results are denoted in the figure by 
circles, whereas the various approximation schemes are represented by the different lines. The inset 
shows the deviations of the various approximation schemes from the QMC data. 
It is clear from the figure that, at least to some extent, all approximation schemes without exception 
provide a fairly accurate description of the equation of state. The spin-wave corrected results however 
provide a better match than the cluster-mean-field approximation which in turn shows only negligible 
improvement over the Gutzwiller-mean-field results. Figure\ \ref{fig:eosTrap2d}(b) shows the density 
in the center of the trap as a function of the characteristic density, a relation that is very useful to 
experimentalists for controlling their density profiles by only changing the total filling 
in their systems once the value of the trapping potential is known.

Figure\ \ref{fig:eosTrap2d} also shows another useful approximation method that is usually applied 
to trapped systems. This is the LDA, which  provides predictions of local 
quantities based on results obtained from matching homogeneous systems. 
Within the LDA, local observables of the confined system are approximated by their values in 
the corresponding homogeneous system, where the chemical potential of the homogeneous case is taken to 
be the corresponding local chemical potential in the trap. In general, the LDA has been shown to give reasonably 
accurate descriptions of other confined systems, but is known to fail in regions where the local 
potential changes rapidly when compared with the correlation length of the relevant phase in the 
homogeneous case. The LDA results here were generated from the QMC results of the homogeneous systems 
studied in the previous section. As the figure illustrates, the LDA provides 
a very good approximation of the QMC results for the density in the center of the trap, although, 
as expected, it underestimates the characteristic density required to form a Mott insulator in 
the center of the trap [Fig.\ \ref{fig:eosTrap2d}(b)].

\subsection{Local densities and compressibilities}

\begin{figure}[!b]
\includegraphics[angle=0,scale=1,width=0.48\textwidth]{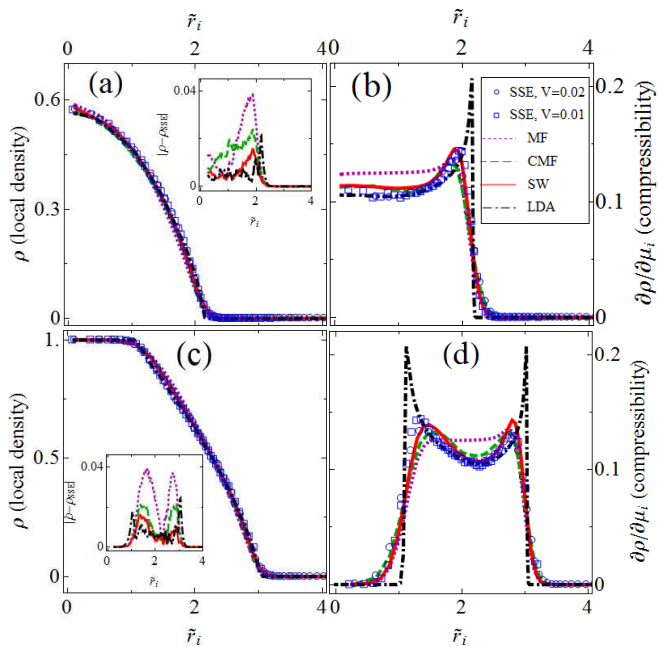}
\caption{\label{fig:densComp} (Color online) (a),(c) Density profiles and (b),(d) local compressibilities
of harmonically trapped hardcore bosons in a two-dimensional lattice as a function of the normalized distance 
$\tilde{r}_i$ from the center of the trap. The top panels correspond to a characteristic density 
$\tilde{\rho} \approx 4.62$ [$\mu/(2dt)=0.175$] for which there is no Mott-insulating region in the center of the trap, 
while the bottom panels correspond to $\tilde{\rho} \approx 15.02$ [$\mu/(2 d t)=1.3$], in which case there is an 
insulating region in the center. The SSE results are indicated by circles ($V=0.01$) and squares ($V=0.02$), 
whereas the dotted, dashed, solid, and dot-dashed lines correspond to the Gutzwiller-mean-field, the cluster-mean-field,
the Gutzwiller-mean-field with spin-wave corrections, and the LDA results, respectively.
The insets in the left panels show the deviations of the approximation schemes from the SSE results.}
\end{figure}

In recent experiments, it has become possible to image local density profiles in optical 
lattices in-situ \cite{gemelke09,bakr09,sherson}. This now allows experimentalists to explore the behavior of local
quantities in a trap. Two quantities that provide insight into the 
properties of the trapped system are the local densities and compressibilities \cite{wessel04}. 
In Fig.\ \ref{fig:densComp}, we show the local radial densities $\rho$ -- averaged over the azimuthal angle --
and the local compressibilities $\partial \rho / \partial \mu_i$ both as a function of $\tilde{r}_i$.
The local compressibility was computed as a numerical derivative of 
the radial density when viewed as a function of the local chemical potential, in much the same way as it would be 
derived in an actual experiment. This is done for two values of chemical 
potential: The top panels correspond to $\tilde{\rho} \approx 4.62$ [$\mu/(2 d t)=0.175$] for which there 
is no Mott-insulating region in the center of the trap, while the bottom panels correspond to 
$\tilde{\rho} \approx 15.02$ [$\mu/(2 d t)=1.3$], in which case there is an insulating region in the 
center, as indicated by the plateau in the density profile when $\tilde{r}$ is small. The QMC data presented 
in the figure correspond to two different values of trap curvature which however trace the same curve, reflecting the 
fact that proper scaling eliminates the dependence of the results on the strength of the trap. 
Interestingly, all approximation schemes are in good agreement with the exact QMC results. 
The spin-wave corrected results and the LDA provide the best match to the QMC results, although
as is evident from the insets of the density profile panels, in the vicinity of the transition between 
the superfluid and the insulating phase, the LDA method shows larger deviations. 
This is also reflected in the compressibilities.  

\subsection{Local phases}

\begin{figure}[!b]
\includegraphics[angle=0,scale=1,width=0.4\textwidth]{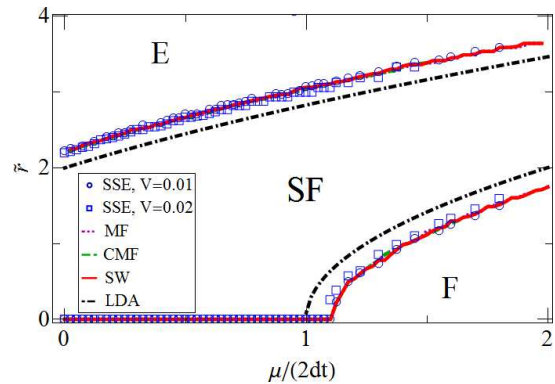}
\caption{\label{fig:sd2d} (Color online) Local phases of harmonically trapped hardcore bosons in a 
two-dimensional lattice; the notations E, SF and F correspond to empty, superfluid, and completely filled lattices,
respectively. The SSE results are denoted by circles ($V=0.02$) and squares ($V=0.01$), 
whereas the various lines are the results of the different approximation schemes: Gutzwiller-mean-field 
(dotted lines), cluster-mean-field (dashed lines), Gutzwiller-mean-field with spin-wave corrections (solid lines),
and LDA (dot-dashed lines). The systems depicted here have $48 \times 48$ lattice sites.
(Note that the Gutzwiller- and cluster-mean-field results here are concealed by the spin-wave-corrected results.)}
\end{figure}

The concept of characteristic density can also provide a general picture of the spatially-separated  
local phases inside the trap as the total filling and/or trapping potential are varied. To show that, we 
have computed the inner and outer radii $\rin$ and $\rout$ which enclose the superfluid region of the bosonic cloud. 
This was done by inspection of the local density profiles obtained via the QMC simulations 
for the different chemical potentials. In practice, the inner radius $\rin$ has been determined 
as the radius inside of which the local density is greater than $0.999$ whereas $\rout$ was defined as 
the radius outside of which the local density is lower than $0.001$. We have verified that our results 
are robust against small deviations from the above values. Figure\ \ref{fig:sd2d} shows the resulting
diagram: Both radii are plotted for two different systems (two different trapping potentials) but with 
the same characteristic density at each point in the diagram. The circles indicate the SSE results with 
$V=0.01$ while squares correspond to $V=0.02$. The lines in the figure indicate the predictions of the 
different approximation schemes. This figure too illustrates that the concept of characteristic density 
is justified even when the LDA is not accurate, as the two systems corresponding to the two values of 
trapping amplitudes trace virtually the same curve. Also, remarkably the various approximation schemes but the LDA 
yield very accurate predictions of both radii. 

Within the LDA, the two radii $\rin$ and $\rout$, at which the local densities are 
$\langle \hat{n}_i \rangle = 1$ and $\langle \hat{n}_i \rangle =0$, respectively,
can be obtained in a straightforward manner: From the homogeneous results discussed in the previous 
sections, we know that these densities are reached at $\mu/(2 d t) =\pm 1$. Translating back to the 
harmonic trap case, this condition becomes $\mu/(2 d t) - \tilde{r}_i^2/(2d)  =\pm 1$. Solving for 
$\tilde{r}_i$ for the two cases, we arrive at:
\beq
&\tilde{r}_{\In, \lda}  & = \textrm{Re} \left[\sqrt{\mu/t -2d}\right] \,,\nonumber\\
&\tilde{r}_{\out, \lda} & = \sqrt{\mu/t +2d} \,.
\eeq
The two radii are indicated by the dot-dashed lines plotted in Fig.\ \ref{fig:sd2d}. As one can see, 
the LDA yields a rather poor prediction of the two radii. 
This is expected however since these radii mark the transition from the superfluid phase to 
the insulating phase in the periodic system where diverging correlations are present. 

\subsection{Thermodynamic properties and the momentum distribution function}

\begin{figure}[!htp]
\includegraphics[angle=0,scale=1,width=0.48\textwidth]{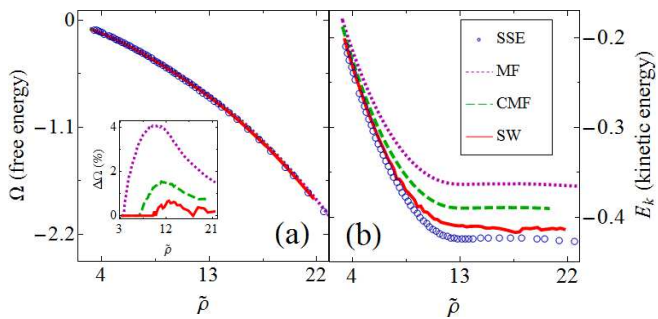}
\caption{\label{fig:alld2trap} (Color online) (a) Free energy and (b) kinetic energy of harmonically 
trapped hardcore bosons in a two-dimensional lattice as a function of the characteristic density. 
The circles indicate the SSE results ($48 \times 48$ sites and $V=0.02$), whereas the dotted, dashed, 
solid, and dot-dashed lines are the mean-field, the cluster-mean-field, the mean-field plus spin-waves, and 
the LDA results, respectively. The inset of the free-energy panel shows the deviation of the approximation 
schemes from the SSE results.}
\end{figure}

To conclude the analysis of the two dimensional case, here we present measurements 
of the kinetic and total (free) energies of the system. These are shown in Fig.\ \ref{fig:alld2trap} 
as a function of the characteristic density. The SSE results are denoted by circles, whereas the dotted, 
dashed and solid lines indicate the Gutzwiller-mean-field, the cluster-mean-field, and the spin-wave-corrected 
approximations. Both the kinetic energy and the inset in the free-energy panel, indicate that 
spin-wave-corrected results provide a far better match than the two mean-field methods.

\begin{figure}[!htp]
\includegraphics[angle=0,scale=1,width=0.48\textwidth]{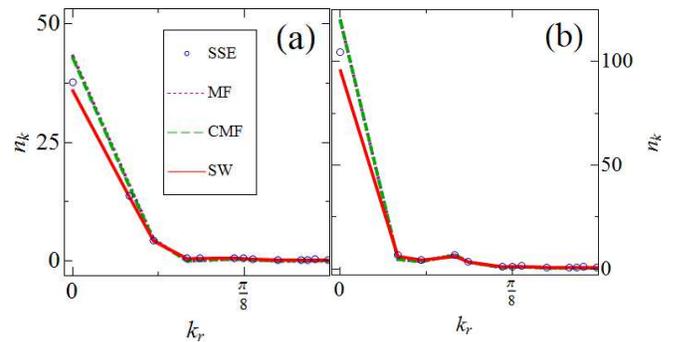}
\caption{\label{fig:mdf2D} (Color online) Momentum distribution of harmonically trapped 
hardcore bosons in a two-dimensional lattice as a function of the radial momentum 
$k_r=\sqrt{k_x^2+k_y^2}$.
(a) $\tilde{\rho} \approx 4.62$ [$\mu/(2dt)=0.175$] and (b) $\tilde{\rho} \approx 15.02$ 
[$\mu/(2dt)=1.3$] for a system with $48 \times 48$ sites. The SSE results are indicated by 
circles whereas the lines are the results of the approximation schemes: Gutzwiller mean-field (dotted lines), 
cluster-mean-field (dashed lines), and Gutzwiller mean-field with spin-wave corrections (solid lines).}
\end{figure}

Next, we analyze the momentum distribution of the model $n(\mathbf{k})$ as a function of
the radial-momentum coordinate $k_r=\sqrt{k_x^2+k_y^2}$ (where, as with the radial densities, 
the azimuthal angle is averaged over). This quantity can be 
directly probed in experiments with ultracold atomic gases via absorption imaging after 
time-of-flight \cite{stoferle04,spielman07,spielman08,greiner02,bloch08}. In the homogeneous case, 
due to the long-range decay of one-particle correlations in the compressible phase, the momentum 
distribution function has a delta peak singularity at $k_r=0$. In general, in the insulating phase, the 
off-diagonal one-particle correlations decay exponentially (they are zero in our hardcore model), 
yielding a broad momentum distribution (completely flat in our case). When a harmonic trap is present, 
the compressible and incompressible insulating phases coexist, so the zero-momentum peak is a 
smoother function of $k_r$. Figure\ \ref{fig:mdf2D} shows the behavior of the momentum distribution
for two values of the characteristic density: $\tilde{\rho} \approx 4.62$ [$\mu/(2 d t)=0.175$], 
for which there is no insulating phase at the center of the trap, and $\tilde{\rho} \approx 15.02$ 
[$\mu/(2dt)=1.3$] for which such a phase exists. Interestingly, it is the latter value for which 
the zero-momentum peak is higher, despite the presence of the insulating phase 
in the center of the trap. This is because the system corresponding to the 
higher characteristic density has a higher number of bosons in the superfluid domain that surrounds the 
Mott insulator. 

Looking at the Gutzwiller- and cluster-mean-field approximations, they both provide rather accurate 
predictions away from zero momentum but overestimate the peak by $\approx 16\%$. The spin-wave corrected 
prediction is much better in both cases, giving only a $\approx 5\%$ error in the worst case.

\section{\label{sec:3dtrap}Trapped hardcore bosons in three dimensions}
In the previous section, we showed that the Gutzwiller-mean-field solution already 
provides a fairly good description of the hardcore Bose-Hubbard model in the presence of a harmonic 
potential, where the deviations from the QMC data were shown to be rather small in most cases.
This was true both for global observables such as the energies shown in Fig.\ \ref{fig:alld2trap},
and also for local observables such as local densities and compressibilities shown in 
Fig.\ \ref{fig:densComp}. When spin-wave corrections were taken into account, the results were 
improved to a point where most errors dropped to below $1\%$, yielding virtually exact 
results.

\begin{figure}[!b]
\includegraphics[angle=0,scale=1,width=0.48\textwidth]{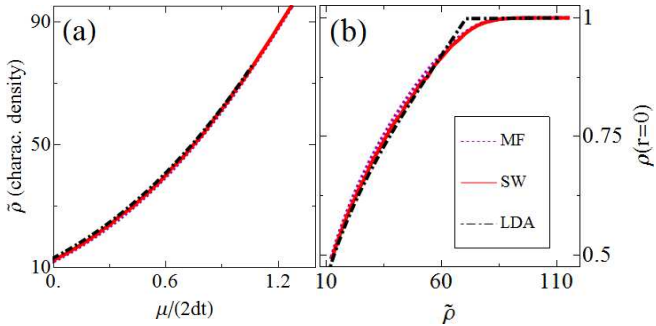}
\caption{\label{fig:eosTrap3d} (Color online) (a) Equation of state and (b) density of bosons at the 
center of the trap for harmonically trapped hardcore bosons in a three-dimensional lattice 
($24 \times 24 \times 24$ sites and $V=0.08$). The dotted lines indicate the Gutzwiller-mean-field while the solid
and dot-dashed lines are the spin-wave-corrected and LDA results, respectively.}
\end{figure}
 
The encouraging results of the previous sections suggest that in three dimensions the spin-wave 
corrected results should be an even better approximation of the exact solution. In light of this, 
in what follows we explore the harmonically trapped three-dimensional hardcore Bose-Hubbard model
using the Gutzwiller-mean-field approach with and without the addition of spin-wave corrections. 
The three-dimensional harmonically-confined system is a good example of a system that becomes 
very demanding computationally if one wants to study the ground state for lattice sizes that are on 
the same order of magnitude as the ones realized experimentally. Having a large linear system size is critical 
when probing local observables that change with distance as is the case of the harmonic trap. 

\begin{figure}[!ht]
\includegraphics[angle=0,scale=1,width=0.4\textwidth]{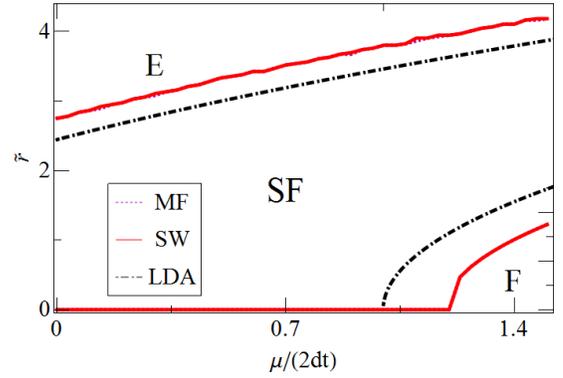}
\caption{\label{fig:sd3d} (Color online) Local phases of harmonically trapped hardcore bosons in 
a three-dimensional lattice; the notations E, SF, and F correspond to empty, superfluid, and completely filled lattices,
respectively. The Gutzwiller-mean-field results are denoted by the dotted 
line, whereas the dashed and solid lines correspond to LDA results and spin-wave corrected results, respectively. 
(Note that the Gutzwiller-mean-field results here are concealed by the spin-wave-corrected results.)}
\end{figure}

\begin{figure}[!ht]
\includegraphics[angle=0,scale=1,width=0.48\textwidth]{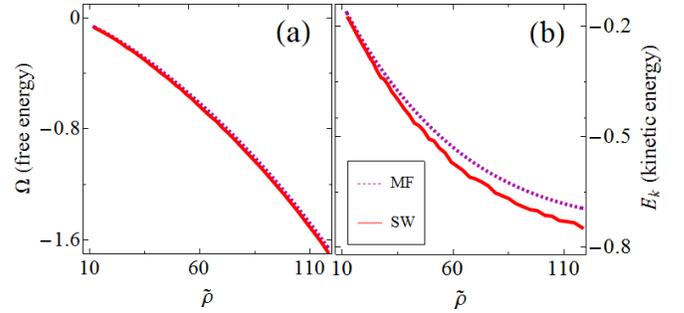}
\caption{\label{fig:alld3trap} (Color online) (a) Free energy and (b) kinetic energy
of harmonically trapped hardcore bosons in a three-dimensional lattice as a function of 
the characteristic density. The dotted lines indicate the Gutzwiller-mean-field results whereas the solid
lines are the spin-wave-corrected results. Here, $V=0.08$.}
\end{figure}

Below we present the results of the three-dimensional analysis: The Gutzwiller-mean-field 
(dotted lines), the spin-wave-corrected (solid lines) and the LDA (dot-dashed lines) predictions of 
the equation of state (Fig.\ \ref{fig:eosTrap3d}), the boundaries of the local phases in the trap 
(Fig.\ \ref{fig:sd3d}) and the energies (Fig.\ \ref{fig:alld3trap}) of harmonically trapped 
hardcore bosons in three dimensions (for a $24 \times 24 \times 24$ lattice). These results can be used 
in current experiments to understand density profiles and other properties of the system in regimes in 
which the local filling does not exceed one and the ratio $U/t$ is very large.

As all three figures indicate, in three dimensions the spin-wave corrections are less pronounced 
than in two dimensions. However, they still provide a discernible improvement over the mean-field results, particularly 
for the kinetic energy. 

\begin{figure}[!htp]
\includegraphics[angle=0,scale=1,width=0.48\textwidth]{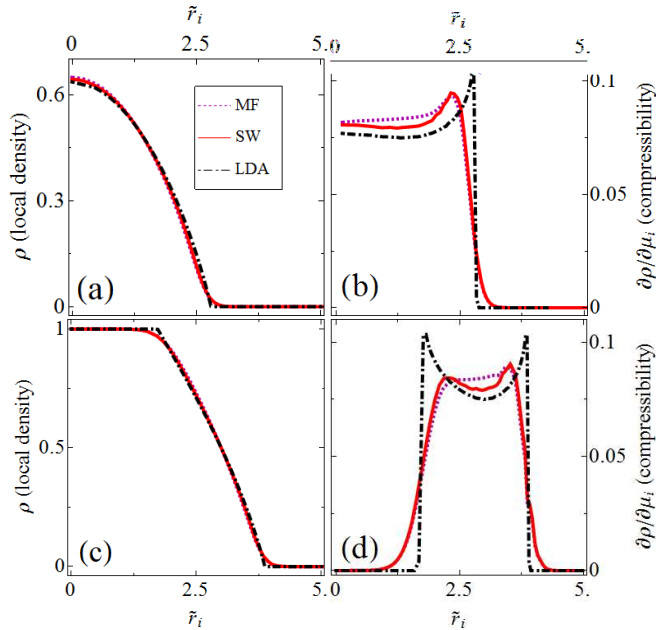}
\caption{\label{fig:densComp3D} (Color online) (a),(c) Density profiles and (b),(d) local compressibilities
of harmonically trapped hardcore bosons in a three-dimensional $24 \times 24 \times 24$ lattice and ($V=0.08$) 
as a function of the normalized distance 
$\tilde{r}_i$ from the center of the trap. The top panels correspond to a characteristic density 
$\tilde{\rho} \approx 23.90$ [$\mu/(2dt)=0.3$] for which there is no Mott-insulating region in the center of the trap, 
while the bottom panels correspond to $\tilde{\rho} \approx 118.12$ [$\mu/(2 d t)=1.5$], in which case there is an 
insulating region in the center. The Gutzwiller-mean-field results are indicated by the dotted lines, 
whereas the dashed, solid, and dot-dashed lines correspond to the Gutzwiller-mean-field, the cluster-mean-field,
the Gutzwiller-mean-field with spin-wave corrections, and the LDA results, respectively.}
\end{figure}

In Fig.\ \ref{fig:densComp3D}, we examine some of the local properties of the three-dimensional system, specifically, 
the radial density profiles $\rho(\tilde{r})$ and the local compressibilities, as they are predicted by the 
Gutzwiller-mean-field, the spin-wave-corrected and the LDA methods. The figure shows the results for two values of chemical 
potential: The top panels correspond to $\tilde{\rho} \approx 23.90$ [$\mu/(2dt)=0.3$] for which there 
is no Mott-insulating region in the center of the trap, while the bottom panels correspond to 
$\tilde{\rho} \approx 118.12$ [$\mu/(2 d t)=1.5$], in which case there is an insulating region in the 
center, as indicated by the plateau in the density profile when $\tilde{r}$ is small. 
As is evident from the figures, in the vicinity of the transition between the superfluid phase and the 
surrounding insulating phases which are rather abrupt, the LDA results detach 
from the mean-field and spin-wave ones. This is also reflected in the compressibilities.  

\section{\label{sec:conc}Summary and conclusions}

In this paper, we have studied the ground-state properties of homogeneous and trapped inhomogeneous systems 
of hardcore bosons in two- and three-dimensional lattices. We have obtained results both by means of quantum 
Monte Carlo (SSE) simulations, and via several analytical approximation schemes, such as the 
Gutzwiller-mean-field approach, a cluster-mean-field method, and a spin-wave analysis which takes quantum 
fluctuations into account. We studied the equation of state and various quantities of interest such as densities 
and compressibilities, condensate and superfluid densities, as well as the momentum distribution function and 
other basic thermodynamic properties.

In the homogeneous case, we showed that the equation of state can be predicted reasonably-well by all approximation
schemes. However, for other quantities such as the kinetic energy, and the superfluid and condensate densities, 
the Gutzwiller and cluster mean-field predictions were found to deviate considerably from the quantum Monte Carlo 
results, particularly close to half filling. Interestingly, the addition of quantum fluctuations through
spin-wave corrections was shown to yield virtually exact results for all quantities but the compressibility. 

In the inhomogeneous case, we found that for the equation of state and some quantities, such as the onsite densities, 
even the simplest method, i.e., the Gutzwiller-mean-field approach, is capable of providing 
a fair analytical description. For other quantities, on the other hand, Gutzwiller-mean-field theory is not a very 
good approximation. For example, for the momentum distribution function we found that it largely overestimates 
the very low momenta occupations. A cluster-mean-field
approach, based on larger cell sizes, did not prove to be much of an improvement over the 
Gutzwiller-mean-field, especially considering the large amount of additional degrees of freedom in the ansatz. 

The addition of spin-wave corrections turned out to be an excellent approximation for all quantities, 
yielding virtually exact results when compared against the QMC data. This, in turn, allowed us to 
study large systems of three-dimensional harmonically-trapped hardcore bosons of sizes that would be very 
demanding for QMC techniques to simulate, with a fraction of the effort. Hence, this analytical approach can
be used to study trapped systems in which the onsite repulsion is much larger than the bandwidth 
and the filling-per-site that is always lower or equal to one.

Finally, we examined the validity of the local density approximation in these harmonically trapped systems. 
We showed that it provides a fairly good description of local quantities as long as the measurement takes place away 
from the superfluid-insulator boundaries, which correspond to a phase transition in the homogeneous case. 

\begin{acknowledgments}
This work was supported by the US Office of Naval Research under Award No.\ N000140910966.
\end{acknowledgments}


\begin{thebibliography}{100}

\bibitem{stoferle04} 
T. St\"oferle, H. Moritz, C. Schori, M. K\"ohl, and T. Esslinger,
Phys. Rev. Lett. {\bf 92}, 130403 (2004).

\bibitem{spielman07} 
I. B. Spielman, W. D. Phillips, and J. V. Porto, 
Phys. Rev. Lett. {\bf 98}, 080404 (2007).

\bibitem{spielman08}
I. B. Spielman, W. D. Phillips, and J. V. Porto, 
Phys. Rev. Lett. {\bf 100}, 120402 (2008).

\bibitem{greiner02} 
M. Greiner, O. Mandel, T. Esslinger, T. W. H\"ansch, and I. Bloch,
Nature (London) {\bf 415}, 39 (2002). 

\bibitem{bloch08} 
I. Bloch, J. Dalibard, and W. Zwerger,
Rev. Mod. Phys. {\bf 80}, 885 (2008).

\bibitem{jimenez10}
K. Jimenez-Garcia, R. L. Compton, Y.-J. Lin, W. D. Phillips, J. V. Porto, 
and I. B. Spielman, arXiv:1003.1541.

\bibitem{trotzky09}
S. Trotzky, L. Pollet, F. Gerbier, U. Schnorrberger, I. Bloch, N.V. Prokof'ev, 
B. Svistunov, and M. Troyer, arXiv:0905.4882.

\bibitem{fisher89}
M. P. A. Fisher, P. B. Weichman, G. Grinstein, and D. S. Fisher,
Phys. Rev. B {\bf 40}, 546 (1989).

\bibitem{batrouni90}
G. G. Batrouni, R. T. Scalettar, and G. T. Zimanyi, 
Phys. Rev. Lett. {\bf 65}, 1765 (1990).

\bibitem{freericks96} 
J. K. Freericks and H. Monien, 
Phys. Rev. B \textbf{53}, 2691 (1996).

\bibitem{kuhner98}
T. D. K\"uhner and H. Monien, 
Phys. Rev. B, {\bf 58}, R14741 (1998).

\bibitem{sansone08}
B. Capogrosso-Sansone, S. G. S\"oyler, N. Prokof'ev, and B. Svistunov,
Phys. Rev. A {\bf 77}, 015602 (2008).

\bibitem{jaksch98}
D. Jaksch, C. Bruder, J. I. Cirac, C. W. Gardiner, and P. Zoller, 
Phys. Rev. Lett. {\bf 81}, 3108 (1998). 

\bibitem{batrouni02}
G. G. Batrouni, V. Rousseau, R. T. Scalettar, M. Rigol, A. Muramatsu, 
P. J. H. Denteneer, and M. Troyer, 
Phys. Rev. Lett. {\bf 89}, 117203 (2002).

\bibitem{kashurnikov02}
V. A. Kashurnikov, N. V. Prokof'ev, and B. V. Svistunov,
Phys. Rev. A {\bf 66}, 031601(R) (2002).

\bibitem{kollath04}
C. Kollath, U. Schollw\"ock, J. von Delft, and W. Zwerger, 
Phys. Rev. A {\bf 69}, 031601(R) (2004).

\bibitem{wessel04}
S. Wessel, F. Alet, M. Troyer, and G. G. Batrouni, 
Phys. Rev. A {\bf 70}, 053615 (2004).

\bibitem{rigol09}
M. Rigol, G. G. Batrouni, V. G. Rousseau, and R. T. Scalettar,
Phys. Rev. A {\bf 79}, 053605 (2009).

\bibitem{1d}
This is with the exception of the special one-dimensional case where the model can be solved 
analytically due to the Jordan-Wigner transformation which enables the mapping of the hardcore 
boson Hamiltonian to that of noninteracting spinless fermions.

\bibitem{ber02}
K. Bernardet, G. G. Batrouni, J.-L. Meunier, G. Schmid, M. Troyer, and A. Dorneich,
Phys. Rev. B {\bf 65}, 104519 (2002).

\bibitem{SSE1}
A. W. Sandvik, 
Phys. Rev. B {\bf 59}, R14157 (1999). 

\bibitem{SSE2}
A. Dorneich and M. Troyer, 
Phys. Rev. E {\bf 64}, 066701 (2001).

\bibitem{hen09}
I. Hen and M. Rigol, 
Phys. Rev. B {\bf 80}, 134508 (2009).

\bibitem{hen10}
I. Hen, M. Iskin, and M. Rigol, 
Phys. Rev. B {\bf 81} 064503 (2010).

\bibitem{rhoS}
M. E. Fisher, M. N. Barber, and D. Jasnow, Phys. Rev. A {\bf 8}, 1111 (1973).

\bibitem{SW1}
K. S. Liu and M. E. Fisher, 
J. Low Temp. Phys. {\bf 10}, 655 (1973).

\bibitem{SW2}
Y.-C. Cheng, 
Phys. Rev. B {\bf 23}, 157 (1981).

\bibitem{SW3}
R. T. Scalettar, G. G. Batrouni, A. P. Kampf, and G. T. Zimanyi, 
Phys. Rev. B {\bf 51}, 8467 (1995).

\bibitem{SW4}
G. Murthy, D. Arovas, and A. Auerbach, 
Phys. Rev. B {\bf 55}, 3104 (1997).

\bibitem{roscilde10}
T. Roscilde, arXiv:1003.4005.

\bibitem{rigol03} 
M. Rigol, A. Muramatsu, G. G. Batrouni, and R. T. Scalettar,
Phys. Rev. Lett. {\bf 91}, 130403 (2003).

\bibitem{rigol04} 
M. Rigol and A. Muramatsu, 
Phys. Rev. A {\bf 69}, 053612 (2004); 
Opt. Commun. {\bf 243}, 33 (2004).

\bibitem{liu05} 
X.-J. Liu, P. D. Drummond, and H. Hu, 
Phys. Rev. Lett. {\bf 94}, 136406 (2005).

\bibitem{campo05} 
V. L. Campo, Jr. and K. Capelle,
Phys. Rev. A {\bf 72}, 061602(R) (2005).

\bibitem{heiselberg06}
H. Heiselberg,
Phys. Rev. A {\bf 74}, 033608 (2006).

\bibitem{xianlong06}
G. Xianlong, M. Polini, M. P. Tosi, V. L. Campo, Jr., K. Capelle, and M. Rigol,
Phys. Rev. B {\bf 73}, 165120 (2006). 

\bibitem{leo08}
L. De Leo, C. Kollath, A. Georges, M. Ferrero, and O. Parcollet,
Phys. Rev. Lett. {\bf 101}, 210403 (2008). 

\bibitem{batrouni08}
G. G. Batrouni, H. R. Krishnamurthy, K. W. Mahmud, V. G. Rousseau, and R. T. Scalettar,
Phys. Rev. A {\bf 78}, 023627 (2008).

\bibitem{camp1}
M. Campostrini and E. Vicari, Phys. Rev. A 81, 023606 (2010).
\bibitem{camp2}
M. Campostrini and E. Vicari, Phys. Rev. A 81, 063614 (2010). 
\bibitem{gemelke09}
N. Gemelke, X. Zhang, C.-L. Hung, and Cheng Chin,
Nature (London) {\bf 460}, 995 (2009).

\bibitem{bakr09}
W. S. Bakr, J. I. Gillen, A. Peng, S. Foelling, and M. Greiner,
Nature (London) {\bf 462}, 74 (2009).
\bibitem{sherson}
J. F. Sherson, C. Weitenberg, M. Endres, M. Cheneau, I. Bloch, and S. Kuhr, Nature (London) 467, 68 (2010).
\end{thebibliography}
\end{document}